# Ehrenfest's paradox for tokamak plasma

Romannikov Alexander

*ITER Domestic Agency - " ITER Project Center ", 123182, pl. Academician Kurchatov, 1, p. 3, Moscow, Russia, a.romannikov @ iterrf.ru*

**Abstract**

A simplified form and some possible theoretical resolutions so-called Ehrenfest's Paradox is described. A relation between physical consequences of this relativistic Paradox and the charge density $\rho$ of a tokamak plasma is shown. Plasma experiments which can resolve the Ehrenfest's Paradox is presented.

**Introduction**

This relativistic Ehrenfest's Paradox was presented in [1] for the first time in 1909. Detailed historical, physical and geometrical description of the Ehrenfest's Paradox can be found in, for example, [2] and references therein. For our experimental purposes, let's present the Ehrenfest's Paradox in the following simplified form. There are two thin rings with radii $R_1$ and $R_2$ (and $R_1 = R_2$ ). The second ring is speed-up to a linear velocity $V$ by some external forces. Let the observers in the laboratory frame measure circumferences of these rings ( $L_1$ and $L_2$ ) and radii in the framework of the relativistic theory methods.

There are basically 3 theoretical hypotheses that present the results of circumference measurements.

The first very old hypothesis we only mention. Concerning of that hypothesis both the radius of rotation ring $R_2$ and the circumference $L_2$ contract by relativistic effect, so that their ratio remains $2\pi \Rightarrow \frac{L_2}{R_2} = 2\pi = \frac{L_1}{R_1}$ [3, 4]. At the contemporary glance the possibility of Ehrenfest's Paradox resolution in frame of this hypothesis is close to zero; see, for example, [2].

The more acceptable resolution of Ehrenfest's Paradox in the laboratory frame for most researchers is:

$$R_1 = R_2 \text{ and } L_1 = 2\pi R_1 = 2\pi R_2 = L_2 \qquad (1).$$

We shall name it "the hypothesis 1". Though it is not so important nuances of "the hypothesis 1" for next part of our investigation, we will consider this hypothesis more in detail. There are two possibilities to obtain that result. Let's introduce the circumference of a rotating ring $L'$ and radius $R_2'$ in the rotating frame (rotating with linear velocity $V$ at the radius $R_2$).

The first and not so widespread approach is based on the assumption like "no Lorentz contraction for rotating frames" [5, 6, 7, 8]. It means that $L_1 = L_2 = L_2'$ and $R_1 = R_2 = R_2'$.

The second and widespread approach is based in the simplified form on next ideas. According to [9], $L' = \dfrac{1}{\sqrt{1 - \dfrac{V^2}{c^2}}} \cdot L_2 = \gamma \cdot L_2$, and $R_2' = R_2$. We shall name it "the condition (1)".

The authors (see, for example, [10, 11, 12, 13, 14, 15]) as a result of the analysis of a metric tensor for a rotating frame, have come to a following conclusion. The condition (1) is fulfilled but the rotating observers see the increase of the circumference in the form of $L' = \gamma \cdot L_1$. The laboratory observers would see a relativistic contraction of a rotating ring in the form of $L_2 = \gamma^{-1} \cdot L' = \gamma^{-1} \gamma \cdot L_1 = L_1$, and $R_1 = R_2' = R_2$.

Let's prolong the previous widespread approach logic, but assume that the measured circumference by the observers in the rotating reference frame does not change due to rotation and is equal to the initial length of the non-rotating ring $L_1$ (see, for example, 16, 17, 18]). Then the laboratory observers would draw a conclusion, that:

$$L_2 = \gamma^{-1} \cdot L_1 \quad \text{and} \quad R_1 = R_2' = R_2 \qquad (2).$$

We shall use the equation (2) latter though only a few researchers support this idea. We shall name it "the hypothesis 2". It is necessary to emphasize, that the geometry of rotation ring points in the laboratory frame is a Non-Euclidean geometry in the case of "the hypothesis 2".

Unfortunately, it is practically impossible to resolve the Ehrenfest's Paradox by observing the real rotating disks or rings. The reason is clear. Centrifugal forces lead to essential deformation of the real rings. Thus, it is impossible to measure very small relativistic effects against the background of that deformation at accessible rotation velocities and size of rings. It is necessary to emphasize that the first investigation which could be conceded the experimental measurement of physical consequences of the Ehrenfest's Paradox was presented only in [18], 2011.

**Physical consequences of the Ehrenfest's paradox for tokamak plasma**

Recently, in [18, 19] the effect of relativistic contraction of an «electron ring» circumference in steady state tokamak plasma rotating in toroidal direction with current velocity $V_e(r)$ has been analyzed. Let $r$ be the minor radius of a tokamak magnetic surface, see Fig.1.

The minor tokamak radius $a$ was assumed to be much less than the major radius $R$, where $\frac{a}{R} \ll 1$, and electron toroidal rotation velocity was assumed to be moderate so that it would be possible to exclude centrifugal forces in the momentum balance of a plasma [20]. The toroidal rotation velocity of «the ion ring» $V_i(r)$, as a rule, is much less than the toroidal rotation velocity of the «electron ring» which is known from experiments [20]. It was supposed that at the initial moment (with no current) plasma is created from neutral gas (hydrogen or deuterium), it is obvious that the electron density $n_e^0(r)$ and the ion density $n_i^0(r)$ are equal, and the difference between the full number of electrons and full number of ions does not vary during a discharge and is equal 0 in tokamak chamber. We can use the last assumption because neutral gas is injected into tokamak chamber and neutral gas is pumped from tokamak. Electrons and ions can move and can be redistributed in the minor radius direction of a tokamak plasma after occurrence of the current.

One can note that the maximum of the experimentally measured radial electric field $E_r(r)$ in tokamak corresponds to the occurrence in plasma a small difference between electron density $n_e(r)$ and ion density $n_i(r)$ in the laboratory frame [18, 19], of the order of $\frac{|n_i(r) - n_e(r)|^{max}}{n_e(r)} \approx (\frac{V_e(r)}{c})^2$; and for the ohmic modes: $\frac{|n_i(r) - n_e(r)|}{n_e(r)} \approx (\frac{V_i(r)}{c})^2$. We shall refer to it as "the condition (2)".

Let us assume at the beginning, that the full number of electrons and ions does not vary during a discharge, all electron density and ion density, ion toroidal rotation velocity and electron rotation velocity are constant and do not depend on the tokamak minor radius $r$. In this case an initial electron density (before plasma current) is $n_e^0$ and ion density is $n_i^0$, where $n_e^0 = n_i^0$.

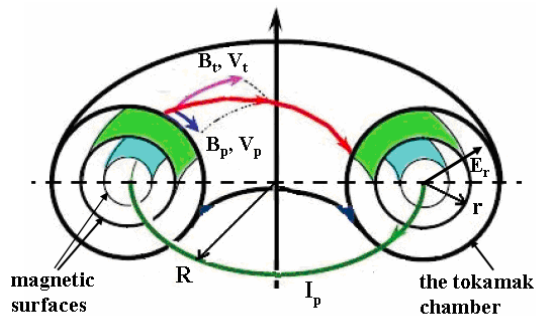

**Fig.1.** A sketch of tokamak.

Therefore, one can say that, actually, there are two thin rings (the electron ring and the ion ring), originally having the same circumference $L^{electron} = 2\pi R = L^{ion}$, which are brought to different toroidal rotation velocities, $V_e$ and $V_i$, where $V_i \ll V_e$. The situation is similar to the one considered above in the context of the Ehrenfest's Paradox. Follow to ideas, summarized by E. L. Feinberg in [21], it is not necessary to investigate particular process of the electron ring and the ion ring acceleration to defined velocities. One can compare only an initial and a final condition. In this case measurement of the part of the electric field which can arise, for example, in the frame of "the hypothesis 2", is much easier, than the investigation of the deformations of a rotating rigid ring. The reason is as follows: on one hand, the current electron velocity can reach hundreds km/s, on the other hand, possible deformations of the "electron ring» due to centrifugal force lead only to the occurrence of dipole components in electric field associated with minor change of radius of the rotating ring. Relativistic contraction of the ring circumference without change of radius and conservation of full electron number (in the frame of "the hypothesis 2") can lead to the occurrence of a monopole component in electric field which is relatively easy to measure, as it will be shown below.

Following [18], let us consider the changing a density of charges $\rho$ in tokamak plasma in frame "the hypothesis 1" and "the hypothesis 2".

"The hypothesis 1".

It is obviously from eq. (1) that tokamak $\rho$ is not changed after appearance of the toroidal rotated electron (current) ring and ion ring in plasma, that is:

$$\rho = 0 \qquad (3).$$

"The hypothesis 2".

Following [18], it is possible to show in this case, that the rotation may create a density of charges $\rho$ in tokamak plasma. If we ignore higher-order terms in $\frac{V^2}{c^2}$ expansion, we can write:

$$\rho = \frac{|e|n_i'}{\sqrt{1-\frac{V_i^2}{c^2}}} - \frac{|e|n_e'}{\sqrt{1-\frac{V_e^2}{c^2}}} \cong |e|n_i' \cdot (1+\frac{1}{2}\frac{V_i^2}{c^2}) - |e|n_e' \cdot (1+\frac{1}{2}\frac{V_e^2}{c^2}) \cong$$

$$\cong -\frac{(|e| \cdot n_e \cdot (V_i - V_e))^2}{2 \cdot c^2 \cdot |e|n_e} + \frac{(|e| \cdot n_e \cdot (V_i - V_e)) \cdot V_i}{c^2} + |e|(n_i' - n_e') = \qquad (4).$$

$$= -\frac{j^2}{2 \cdot c^2 \cdot |e|n_e} + \frac{j \cdot V_i}{c^2}$$

Where $n_e'$ and $n_i'$ are: the electron density in the rotating frame with the velocity $V_e$ and the ion density in the rotating frame with the velocity $V_i$. We have took into account in frame of "the hypothesis 2" - $L_{rot}^{electron} = L^{electron} \cdot \sqrt{1-\frac{V_e^2}{c^2}} \neq L_{rot}^{ion} = L^{ion} \cdot \sqrt{1-\frac{V_i^2}{c^2}}$ and $n_i' = n_i^0 = n_e^0 = n_e'$.

Change of the charge density in this case is only associated with relativistic change of the denominator in the expression for density. Let us note that $\rho$ depends on parameters measured in the laboratory frame: the current density $j$, the electron density $n_e$ and the ion toroidal rotation velocity $V_i$.

So we have calculated the charge density $\rho$ in the each point inside a tokamak non moving chamber in laboratory frame. One can "forget" about the particular nature of that charge density $\rho$ related to Non-Euclidean geometry of rotation electron (ion) ring points of the tokamak plasma in the laboratory frame and one can use the Poisson equation with $\rho$ taken from eq. (4) to calculate the electrostatic radial electric field in the tokamak plasma. Hence, $E_r(r)$ in that simple tokamak plasma is created by two relativistic terms in the density of charges $\rho$, eq. (4), appeared in the laboratory frame.

In case of a real tokamak, plasma parameters depend on minor radius of magnetic surfaces. Consideration of such dependence for the purpose of calculation of tokamak plasma charge density is shown in [18, 19] in detail for $\frac{a}{R} \ll 1$. The principal point here is the consideration of each nested magnetic surface with plasma just in a thin hollow plasma ring.

Following [18], we can rewrite the eq. (3) in case of "the hypothesis 1" so as to take into account the processes of redistribution of electrons and ions on the minor radius by plasma diffusion (convection):

$$\rho(r) \cong |e|(n_i^d(r) - n_e^d(r)) \qquad (5).$$

Due to the electron and ion diffusion (change of the numerator in the expression for density) additional volume charge densities in plasma can arise, and it can be expressed by the term $|e|(n_i^d(r) - n_e^d(r))$ in (5). Hence, only the diffusion (convection) of ions and electrons could be created $E_r(r)$ in tokamak plasma.

In the case of "the hypothesis 2", we can rewrite the eq. (4) in the following form:

$$\rho(r) \cong -\frac{j^2(r)}{2 \cdot c^2 \cdot |e| n_e(r)} + \frac{j(r) \cdot V_i(r)}{c^2} + |e|(n_i^d(r) - n_e^d(r)) \qquad (6),$$

where $j = |e| \cdot n_e(r) \cdot (V_i(r) - V_e(r))$.

The eq. (6) has two relativistic terms, and, by the way, the "condition (2)" is not casual coincidence in this case. Let us emphasize again that $\rho(r)$ depends on the plasma parameters measured in laboratory frame: the current density $j(r)$, the electron density $n_e(r)$, the ion toroidal rotation velocity $V_i(r)$ and the diffusion (convection) term. So we have calculated the charge density $\rho$ in the each point inside tokamak non-moving chamber in the laboratory frame, eq. (6). As emphasized above one can "forget" about the particular nature of that charge density $\rho$ related partially to Non-Euclidean geometry of rotation electron (ion) ring points of the tokamak plasma in the laboratory frame and one can use the Poisson equation with $\rho$ taken from eq. (6) to calculate the electrostatic radial electric field in the tokamak plasma. In our consideration the diffusion (convection) term is not determined. We can mention about one integral property of the diffusion (convention) term, which is a consequence of the physical assumption that the difference between the full number of electrons and full number of ions does not vary during a discharge and is equal 0 in tokamak chamber. It is:

$$\int_{V_{ch}} |e|(n_i^d(r) - n_e^d(r)) dV = 0 \qquad (7),$$

where $V_{ch}$ is the volume of the toroidal tokamak chamber. The diffusion (convention) term can be determined in the frame of different approaches, see [19].

**T11M tokamak experiment**

Having accepted "the hypothesis 2", we have seen that plasma current creates relativistic volume charge density equal to $-\frac{j^2}{2 \cdot c^2 \cdot |e| n_e}$. The second relativistic right-hand term of eq. (6)

for plasma usually is more than five times smaller than $-\dfrac{j^2}{2 \cdot c^2 \cdot |e| n_e}$. The third term is the symmetrical redistribution of charges by the diffusion (convection) along the minor radius in a plasma chamber. Thus $-\dfrac{j^2}{2 \cdot c^2 \cdot |e| n_e}$ can be crucial in the creation of $E_r(r)$, especially at the beginning of a discharge, and if plasma has modulated current. For a tokamak plasma contained in metallic chamber, $E_r(r)$ can modify the chamber electric potential

$$\Delta\varphi \cong \dfrac{\int_{V_{ch}} \rho(r) dV}{C} \approx \dfrac{\int_{V_{ch}}(-\dfrac{j^2(r)}{2 \cdot c^2 \cdot |e| n_e(r)} + |e|(n_i^d(r) - n_e^d(r)))dV}{C} = \dfrac{\int_{V_{ch}}(-\dfrac{j^2(r)}{2 \cdot c^2 \cdot |e| n_e(r)})dV}{C}$$

with respect to the ground; see the eq. (7). Chamber electric potential is proportional in this case to the volume of plasma, the averaged value of $-\dfrac{j^2}{2 \cdot c^2 \cdot |e| n_e}$, the electric capacitance of close metallic tokamak chamber C and relates to chamber RC time.

If one wants to measure the potential of the tokamak chamber Δφ(t) during the discharge, one can expect two options. In the case of "the hypothesis 1" - no change of Δφ(t) due to the plasma current will occur, i.e. Δφ(t) = 0, see the eq. (7); in the case of "the hypothesis 2" - the potential of the chamber will change proportional to $\dfrac{\int_{V_{ch}}(-\dfrac{j^2(r)}{2 \cdot c^2 \cdot |e| n_e(r)})dV}{C}$.

Thus, measurement of tokamak chamber potential Δφ(t) during discharges could resolve the Ehrenfest's Paradox in principle.

The first series of special experiments for electric potential measurements in several tokamak chamber points was carried out at T-11M tokamak (main plasma parameters in presented shots were: deuterium plasma, the average steady-state electron density <$n_e$> ~$10^{13}$ cm$^{-3}$, the plasma current Ip ~50 kA, $r$ =20 cm, $R$ =70 cm) with modulated current [22, 18]. The example of the typical measurement is shown on Fig.2. For the purpose of calculation of the theoretical dependence (triangles and dashed curve on Fig. 2) we have used: a) experimental data for plasma current and electron density; b) experimental chamber resistivity R = ~4 MOm; c) experimental chamber RC ~ 2.5 ms. Electron density diagnostics did not give us adequate information for few milliseconds in the beginning of discharge. We have extrapolated the electron density growth during the first ~8 ms by a linear function.

One can see satisfactory coincidence of theoretical calculation results based on "the hypothesis 2"with the experimental results.

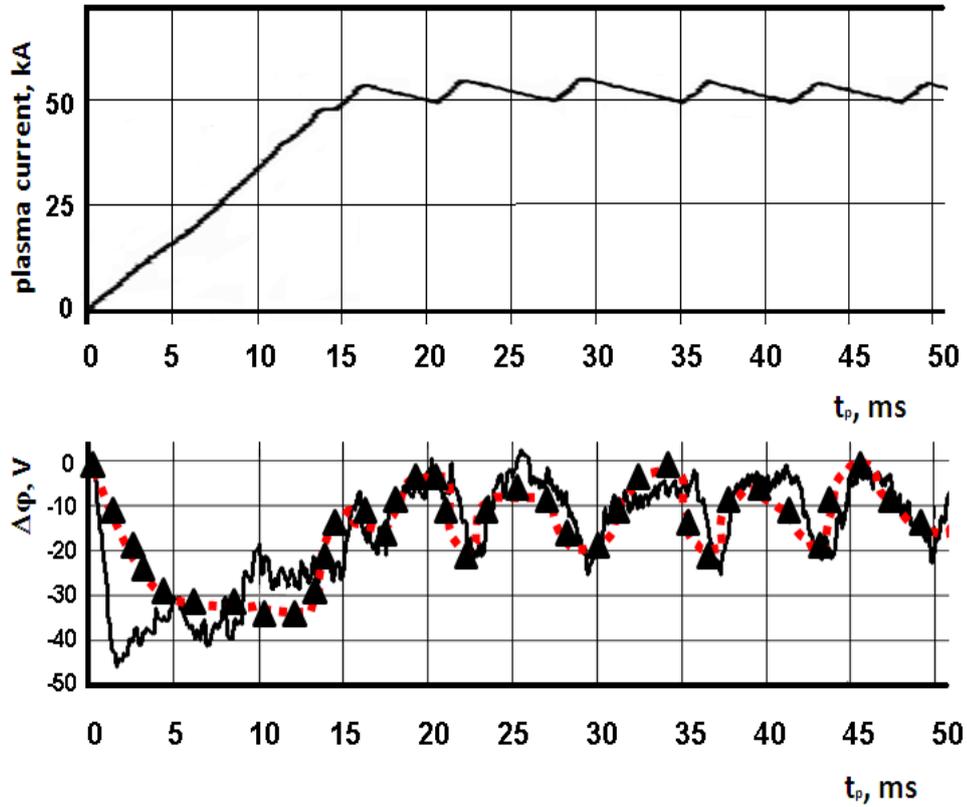

**Fig.2.** Time dependence of plasma current $I_p(t)$ and the tokamak chamber electric potential $\Delta\varphi(t)$ during the discharge, T-11M tokamak [22]. Solid curve is the experimental dependence of $\Delta\varphi(t)$; triangles and dashed curve are the theoretical dependence.

**Physical consequences of equation (7) and some tokamak experiments**

Radial electric field $E_r(r)$ plays an important role in various modes of improved plasma confinement in tokamak [23]. Some of those modes will be used in the thermonuclear reactor ITER [24] which is currently under construction.

Because "the condition 2" exists in tokamak plasma, we usually cannot use the Poisson equation for calculation of $E_r(r)$ - it is not possible to measure or even to calculate independently $n_e(r)$ and $n_i(r)$ in the laboratory frame with necessary accuracy.

Unfortunately, another approach - the ambipolarity equation for radial flows, cannot be used also since ambipolarity emerges automatically from toroidal symmetry of the considered configuration [25, 26]. For those reasons more complex approaches to estimation of $E_r(r)$ are used. There are many successful methods for calculation of $E_r(r)$, see, for example, [20, 27] and references within [20, 27]. It is necessary to emphasize there are some experimental results concerning $E_r(r)$, see, for example, [28, 29, 30, 31, 32], which are difficult to explain within a simple and single approach. Those experimental results can be explained not only by complex contemporary theories but enough simple approach based on eq. (6), too. It will be shown below.

Let us assume that the charge density, eq. (6), is created $E_r(r)$ in a tokamak plasma. To compare the results of this approach to $E_r(r)$ with the results of some actual experiments, we take into account the additional plasma equation, see, for example, [20] (which is derived from radial equilibrium of forces on a magnetic surface):

$$E_r(r) \cong \frac{V_t(r) \cdot B_p(r)}{c} + \frac{1}{|e| \cdot n_i(r)} \cdot \frac{dP_i(r)}{dr} - \frac{V_p(r) \cdot B_t}{c} \qquad (8),$$

where $V_p(r)$ and $V_t(r)$ ($V_t(r) \equiv V_i(r)$ in the article) are the velocities of the poloidal and toroidal rotation of plasma ions and, hence, of the plasma as a whole (the velocities are low enough so the centrifugal effect may be omitted); $c$ is the speed of light, $e$ is the electron charge; $n_i(r)$ and $P_i(r)$ are the density and pressure of plasma ions; $B_p(r)$ and $B_t(r)$ are poloidal and toroidal magnetic fields. To establish main features of relations between $E_r(r)$ and $V_t(r)$, as well as other plasma parameters, we may ignore the weak poloidal dependence of parameters in eq. (8) ($1 \pm \frac{r}{R} \cdot \cos\vartheta$, where $R$ is the major radius of the tokamak, $\vartheta$ is the poloidal angle; $\frac{r}{R} \ll 1$). A relation of type (8) is always true when the plasma is in the steady state (only these states are considered below). For the sake of simplicity let's assume that magnetic surfaces are nested cylinders with small radii $r$ and plasma consists of electrons and, for example, deuterium ions. Let's also assume that velocity of poloidal rotation may be taken from experiments or derived from neo-classical theory.

We can express $E_r(r)$ and $V_i(r)$ independently using eq. (6) and eq. (8):

$$E_r(r) \cong -\frac{B_p(r)}{c} \cdot \int_r^a \left[ V_p(\xi) \cdot \frac{B_t}{B_p(\xi)} - \frac{c}{|e| \cdot n_e(\xi) \cdot B_p(\xi)} \cdot \frac{dP_i(\xi)}{d\xi} - \frac{j(\xi)}{2 \cdot |e| n_e(\xi)} + |e|(n_i^d(\xi) - n_e^d(\xi)) \cdot \frac{c^2}{j(\xi)} \right] \times$$
$$\times \left( \frac{1}{\xi} + \frac{1}{B_p(\xi)} \frac{dB_p(\xi)}{d\xi} \right) d\xi + \left[ \frac{V_p(r) \cdot B_t}{c} - \frac{1}{|e| \cdot n_e(r)} \cdot \frac{dP_i(r)}{dr} \right]_{r=a} \quad (9),$$

$$V_i(r) \cong V_p(r) \frac{B_t}{B_p(r)} - \frac{c}{|e| \cdot n_e(r) \cdot B_p(r)} \cdot \frac{dP_i(r)}{dr} - \left[ V_p(r) \frac{B_t}{B_p(r)} - \frac{c}{|e| \cdot n_e(r) \cdot B_p(r)} \cdot \frac{dP_i(r)}{dr} \right]_{r=a} -$$
$$- \int_r^a \left[ V_p(\xi) \cdot \frac{B_t}{B_p(\xi)} - \frac{c}{|e| \cdot n_e(\xi) \cdot B_p(\xi)} \cdot \frac{dP_i(\xi)}{d\xi} - \frac{j(\xi)}{2 \cdot |e| n_e(\xi)} + |e|(n_i^d(\xi) - n_e^d(\xi)) \cdot \frac{c^2}{j(\xi)} \right] \times \quad (10).$$
$$\times \left( \frac{1}{\xi} + \frac{1}{B_p(\xi)} \frac{dB_p(\xi)}{d\xi} \right) d\xi$$

We take into account the assumption that the toroidal rotation velocity at the plasma boundary with $r = a$ is close to zero [33].

Below we give important quantitative and qualitative correlations of theoretical results (using eqs. (6, 9, 10)) with real tokamak experimental data [19, 28, 29, 30, 31, 32, 34]. The first important conclusion from eq. (10): if $-\frac{j^2(r)}{2 \cdot c^2 \cdot |e| n_e(r)} + |e|(n_i^d(r) - n_e^d(r))$ is not close to zero in plasma core then a toroidal ion beam is created in that region (its velocity logarithmically tending to infinity [19, 22]). So we have to suppose that $-\frac{j^2(r)}{2 \cdot c^2 \cdot |e| n_e(r)} + |e|(n_i^d(r) - n_e^d(r)) \cong 0$ at the core of the plasma.

This leads to following consequences. One can rewrite in simplified form the eq. (9) [19]:

$$E_r(r) \cong \frac{B_p(r) \cdot V_i(r)}{c} - \frac{1}{r \cdot c} \int_0^r B_p(\xi) \cdot \frac{dV_i(\xi)}{d\xi} \xi \cdot d\xi \quad (11),$$

where $B_p(r) \cong \frac{4\pi}{r \cdot c} \int_0^r j(\xi) \cdot \xi \cdot d\xi$.

Calculated $E_r(r)$ using eq. (11) is presented on Fig.3 for typical ohmic discharge in TCV tokamak [28]. One can see satisfactory coincidence of theoretical calculation results based on eq.(11) with the experimental results.

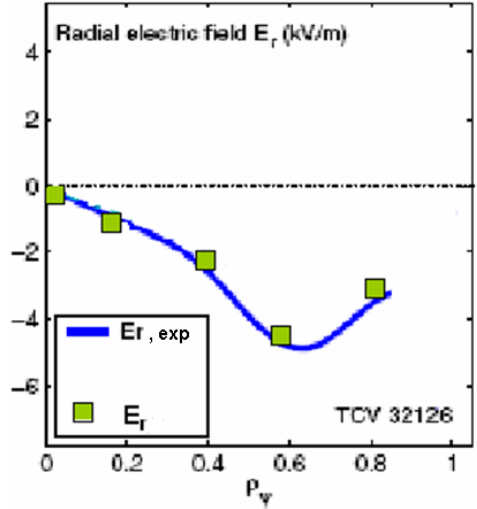

**Fig.3.** Example of radial electric field $E_r(r)$ profile calculation in a typical ohmic discharge in TCV tokamak [28].

$\rho_\psi$ is the effective radius of tokamak magnetic surface. Solid curve is the experimental $E_r(r)$ profile; squares are the $E_r(r)$ values calculated with eq. (9) using the experimental toroidal rotation plasma velocity profile.

Investigation of so-called «locked» mode is another quantitative example of this approach. During that mode plasma stops rotating on the toroidal direction on all magnetic surfaces and experimental $E_r(r)$ becomes close to zero everywhere [29]. $E_r(r) \cong 0$, if $V_i(r) \cong 0$ is enough difficult to explain by, for example, approaches based only on eq. (8). But it is the trivial consequence of eq. (11).

Sign and typical value of toroidal rotation velocitiy in plasma core in most ohmic modes and some modes with ICRH (if ion toroidal rotation velocity is opposite to the plasma current and is equal to 10÷100 km/s [20, 31]) are correctly described by eq. (10) with

$$-\frac{j^2(r)}{2 \cdot c^2 \cdot |e| n_e(r)} + |e|(n_i^d(r) - n_e^d(r)) \cong 0$$ for real ion pressure profiles [19, 22].

If the sum of $-\frac{j^2(r)}{2 \cdot c^2 \cdot |e| n_e(r)} + |e|(n_i^d(r) - n_e^d(r))$ in eq. (10) becomes less than zero at plasma periphery (this often indicates suppression of electron loss at plasma periphery) then plasma core begin rotating on the current direction (which correlates with experimental data [30, 19]).

An important consequence of eq. (9) and eq. (10) is the fact that plasma confinement is better in mode with co-current plasma rotation and positive $E_r(r)$ than in mode with counter-current plasma rotation and negative $E_r(r)$ [19] (if other plasma parameters are similar and two modes differ from each other only in direction of rotation and $E_r(r)$ sign). This is confirmed by experimental data [32].

The integral relation between plasma parameters, eq. (9) and eq. (10), leads us to following fact: a local variation of plasma parameters in some region, for example, on some periphery magnetic surface, leads to «instantaneous» total change of toroidal rotation plasma velocity and $E_r(r)$ on whole magnetic surfaces inside the perturbated magnetic surface. Such non-diffusive penetration of perturbations was observed in experiments; see, for example, [34].

**Conclusion**

1. Presented relativistic theory of radial electric field formation, based on eq. (6), can explain, sometimes quantitatively and more often qualitatively, a lot of experimental tokamak results where $E_r(r)$ and $V_i(r)$ are measured. Some particular examples are presented in the article.

2. Tokamak plasma can be a tool for the research of possible physical consequences of the Ehrenfest's Paradox. Measurement of tokamak chamber potential Δφ(t) with respect to the ground during discharges could resolve that Paradox in principle. The plasma created from initially neutral gas inside metallic tokamak chamber can affect to Δφ(t) by two following ways.

    a) The most expected effect is Δφ(t)=0. In this case the Ehrenfest's Paradox should be resolved in the frame of "the hypothesis 1".

    b) The particular effect is $\Delta\varphi(t) \approx \dfrac{\int_{V_{ch}}(-\dfrac{j^2(r)}{2\cdot c^2 \cdot |e| n_e(r)})dV}{C}$. In this case the Ehrenfest's Paradox should be resolved in the frame of "the hypothesis 2".

3. Available experiments of Δφ(t) measurements described above and good agreement of theoretical results, based on eq. (6), with experimental results, see the first item of Conclusion, show, that the Ehrenfest's Paradox could be resolved in the frame of "the hypothesis 2" much more probably than in the frame of "the hypothesis 1". It has the important consequence. The geometry of rotation electron (ion) ring points of the tokamak plasma in the laboratory frame should be a Non-Euclidean geometry.

Enlightening discussions with Yury Romannikov are gratefully acknowledged.